\documentclass{wecapp}
\usepackage{graphicx}

\usepackage{natbib}

\begin{document}

\title{WeCAPP - The Wendelstein Calar Alto Pixellensing Project I}

\subtitle{Tracing Dark and Bright Matter in \object{M\,31}\thanks{Based 
on observations obtained at the Wendelstein Ob\-ser\-va\-tory of the 
University Observatory Munich.}}

\author{A.~Riffeser\thanks{Visiting astronomer at the German-Spanish 
Astronomical Center, Calar Alto, operated by the Max-Planck-Institut f\"ur 
Astronomie, Heidelberg, jointly with the Spanish National Commission for 
Astronomy.}
\and  J.~Fliri$^{\star\star}$ \and C.~A.~G\"ossl \and  
R.~Bender \and U.~Hopp \and O.~B\"arnbantner \and  C.~Ries \and \\
H.~Barwig \and  S.~Seitz \and W.~Mitsch}

\institute{Universit\"ats-Sternwarte M\"unchen, Scheinerstr. 1, 
  D-81679 M\"unchen, Germany}

\offprints{A. Riffeser,\\ \email{arri@usm.uni-muenchen.de}}

\date{Received 12 April 2001 / Accepted 6 September 2001}

\abstract{
We present WeCAPP, a long term monitoring project searching for microlensing
events towards M\,31. Since 1997 the bulge of M\,31 was monitored in two
different wavebands with the Wendelstein 0.8\,m telescope. In 1999 we 
extended our observations to the Calar Alto 1.23\,m telescope. Observing 
simultaneously at these two sites we obtained a time coverage of 53\,\%
during the observability of M\,31.
To check thousands of frames for variability of unresolved sources, we 
used the optimal image subtraction method (OIS) by 
\citet{1998ApJ...503..325A}. 
This enabled us to minimize the residuals in the difference image
analysis (DIA) and to detect variable sources with amplitudes at the photon 
noise level. Thus we can detect microlensing events 
with corresponding amplifications $A>10$ of red clump giants 
with $M_\mathrm{I}=0$.
}

\maketitle

\section{Introduction} 

In the last decade microlensing studies proved to be a powerful tool
for searching baryonic dark matter in the Galactic halo.
\\
Several groups like the MACHO collaboration \citep{1993Natur.365..621A},
OGLE \citep{1993AcA....43..289U}, EROS \citep{1993Natur.365..623A} 
and DUO \citep{1995Msngr..80...31A} followed the suggestion of 
\citet{1986ApJ...304....1P} and surveyed millions 
of stars in the Large and Small Magellanic Clouds (LMC, SMC) and in 
the Galactic bulge for variability induced by gravitational microlensing.
Although all of them discovered events compatible with gravitational 
lensing by MACHOs (Massive Astrophysical Compact Halo Objects) 
\citep{1994ApJ...435L.113P,1996A&A...314...94A,1997ApJ...486..697A,
1997A&A...326....1A,1998A&A...332....1P,1999A&A...343...10A,
1999A&A...351...87E,2000ApJ...542..281A,2000ApJ...541..734A,
2000AcA....50....1U}
they were not able to derive unambiguous constraints on the amount of baryonic
dark matter and its distribution in the Galactic halo 
\citep[e.g.][and references therein]{2000A&A...355L..39L,2000ApJ...529..917E}.
\\
\citet{1992ApJ...399L..43C} and \citet{1993A&A...277....1B}
suggested to include M\,31 in future lensing 
surveys and pointed out that it should be an ideal target for these
kind of experiments. In contrast to microlensing studies towards the LMC
and the SMC, which are restricted to similar lines of sight through the 
Galactic halo, one can study many different lines of sight to M\,31, 
which allow to separate between self-lensing and true MACHO events.
\\
Since the optical depth for Galactic MACHOs is much greater 
towards M\,31 than towards the LMC, SMC or the Galactic bulge one expects
event rates greater than in previous lensing studies. Furthermore M\,31 
contributes an additional MACHO population as it possesses a dark halo 
of its own. Thus, three populations may contribute to the optical depth
along the line of sight: MACHOs in the Galactic halo, MACHOs in the halo of 
M\,31 and finally stars in the bulge and the disk of M\,31 itself, a 
contribution dubbed self-lensing.
\\
The high inclination of M\,31 ($77^{\circ}$) \citep{1987A&AS...69..311W}
produces a near-far asymmetry of the event rates. The near side of the 
M\,31 disk will show less events than the more distant one 
\citep{1992ApJ...399L..43C}. Since Galactic halo-lensing as well as 
self-lensing events will not show this feature, a detected asymmetry 
will be an unambiguous proof for the existence of M\,31 MACHOs.
\\
As most of the sources for possible lensing events are not resolved
at M\,31's distance of 770 kpc \citep{1990ApJ...365..186F} the name 
`pixellensing' \citep{1996ApJ...470..201G}
was adopted for these kind of microlensing studies. 
In the mid nineties two projects started pixellensing surveys towards 
M\,31, AGAPE \citep{1997A&A...324..843A} and Columbia/VATT 
\citep{1996AJ....112.2872T}. First candidate events were reported 
\citep{1999A&A...344L..49A,1996ApJ...473L..87C} but could 
not yet be confirmed as MACHOs. This was partly due to an insufficient 
time coverage which did not rule out variable stars as possible sources.
\\ 
1999 two new pixellensing projects, POINT-AGAPE \citep{KERPOI},
who reported recently a first candidate microlensing event 
\citep{2001ApJ...553L.137A}, and MEGA \citep{1999ASP..182..409C},
the successor of Columbia/VATT, began their systematical observations 
of M\,31. 
Another project, SLOTT-AGAPE \citep{SLOTT}  will join them this year.
\\
The Wendelstein Calar Alto Pixellensing Project started with a test 
and preparation phase on Wendelstein as WePP in
autumn 1997 before it graduated after two campaigns to WeCAPP in summer
1999 by using two sites for the survey. 
\\
In this paper we will give an introduction to the project
including information about the data obtained in three years 
and our reduction pipeline.  In Sect.~2 we briefly discuss the 
basic principles of pixellensing. In Sect.~3 we will
give an overview of the project including information 
about the sites used and the data obtained during WeCAPP. 
Section~4 refers to our data reduction pipeline and describes
how light curves are extracted. In Sect.~5 we show first light curves 
and Sect.~6 summarizes the paper.

\section{Pixellensing}

In microlensing surveys of uncrowded fields a resolved star is 
amplified by a function $A$ which can be measured 
from the light curve and which yields direct information of the lensing 
parameters 
\citep{1986ApJ...304....1P}:
\begin{equation} 
  A(u)= \frac{u^2+2}{u\sqrt{u^2+4}} \;;\;
  u^2=\mu_\mathrm{E}^2 (t-t_0)^2+u_0^2 
  \label{classic2}
\end{equation}
with $\mu_\mathrm{E}=1/t_\mathrm{E}$, the inverse Einstein ring crossing 
time, $u_0$ the angular impact parameter in units of the angular Einstein 
ring radius $\theta_\mathrm{E}$ and $t_0$ the time of maximum amplification.
\\
The angular Einstein ring radius $\theta_\mathrm{E}$ is connected with 
the physical Einstein ring radius $r_\mathrm{E}$ and the properties of the 
gravitational lens by
\begin{equation} 
  \theta_\mathrm{E}=\frac{r_\mathrm{E}}{D_\mathrm{l}}  \;;\;  r_\mathrm{E}=
  \bigg(\frac{4GM}{c^2}\frac{D_\mathrm{l}D_\mathrm{ls}}{D_\mathrm{s}}\bigg)
  ^\frac{1}{2}
\end{equation}
with the mass of the lens $M$, and the distances between the observer
and lens  $D_\mathrm{l}$, observer and source $D_\mathrm{s}$, 
and lens and source $D_\mathrm{ls}$. 
According to Eq. \ref{classic2} a lensing event will produce a 
symmetric and, as lensing does not depend on the colour of the source,
achromatic light curve (see also Sect.~3).
\\
In more crowded fields it is not possible to determine 
the amplification $A$ unambiguously because many 
unresolved sources may lie inside the solid angle of the point spread 
function $\Omega_\mathrm{PSF}$ of a bright source. If one of these 
unresolved sources is amplified, the event could erroneously
be attributed to the bright star, which results in a strong amplification bias
\citep{1997ApJ...484..555H,1997A&A...321..424A,1997ApJ...487...55W,
1998ApJ...498..156G}. 
\\
Similarly, in pixellensing studies the true amplification cannot be 
determined because many stars fall in one resolution element.
\\
Nevertheless one can still 
construct a light curve $\Delta F(t)$ by subtracting a reference frame 
from the image, in which a lensing event takes place:
\begin{equation}
\Delta F(t)=F(t)-B=F_i \cdot [A(t)-1] \;;\; B=F_i+B_{\mathrm{res}}
\label{classic1}
\end{equation}
with $F_i$ the flux of the source after or before lensing and 
$B_\mathrm{res}=\sum_{j \ne i} F_j$ being the residual flux from 
unlensed sources inside 
$\Omega_\mathrm{PSF}$. As there are many sources contributing to the 
flux inside $\Omega_\mathrm{PSF}$, pixellensing events will only
be detected if the amplification $A$  or the intrinsic flux $F_i$ of 
the lensed source are high. However, this difficulty in identifying events 
is balanced by the large number of possible sources for lensing events in 
the bulge of M\,31. 
\\
As pointed out by \citet{1996ApJ...470..201G}, the main difference between 
classical microlensing and pixellensing consists in the fact, that
in the latter case the noise within $\Omega_\mathrm{PSF}$ is dominated 
by unlensed sources and therefore stays virtually constant during an event. 
As only events with a high amplification 
can be detected, the Einstein timescale $t_\mathrm{E}$ is not a general 
observable in pixellensing. Therefore $t_{\mathrm{FWHM}}$ which 
describes the width of a lensing light curve at half of its 
maximum value is the only timescale one is able to measure with a certain
accuracy. Based on a previous work of \citet{1999ApJ...510L..29G}, 
who pointed out that the optical depth towards M\,31 can be estimated 
without knowing $t_\mathrm{E}$, \citet{2000ApJ...530..578B} showed 
how the measurement of a moment of the light curve $t_{\sigma n}$ 
permits to calculate the Einstein time $t_\mathrm{E}$ for a particular 
lensing event. Furthermore principal component analysis of the light
curves can yield a less biased information about the mass function of the 
lenses as shown by \citet{2001MNRAS.320..341A}.
\\
\citet{1996ApJ...472..108H} has estimated the amount of expected microlensing 
events with a cumulative event signal-to-noise ratio
$(S/N)_\mathrm{min}=20$  
\citep[for a further description see][]{1996ApJ...470..201G}. 
The assumption he used was that the whole dark matter halo consists of MACHOs.
The calculations are mainly depending on the luminosity function of the M\,31 
stars and on the spatial distribution of the lenses. From Han's paper we 
derive the following scaling relation for the event rate:
\begin{eqnarray}
  \Gamma &=& 94\,\mathrm{ev/yr}\;
  {\left(\frac{t_\mathrm{camp}}{1/3\,\mathrm{yr}}\right)}
  {\left(\frac{t_\mathrm{cyc}}{1\,\mathrm{d}}\right)}^{-1}
  {\left(\frac{t_\mathrm{exp}}{4\,\mathrm{h}}\right)} 
  {\left(\frac{\epsilon}{12\,\mathrm{phot/s}}\right)}\cdot \nonumber\\
  & & \cdot
  {\left(\frac{\theta_\mathrm{see}}{1\,\arcsec}\right)}^{-2}
  {\left(\frac{\eta_\mathrm{sky}^2+\eta_\mathrm{psf}^2}{3.35}\right)}^{-1}
  {\left(\frac{\Omega_\mathrm{ccd}}{70\,\mathrm{arcmin}^2}\right)}
\end{eqnarray}
For the typical parameters of our 1999/2000 campaign this  results in a
predicted event rate $\Gamma=27\,\mathrm{ev/yr}$, with
the length of the campaign $t_\mathrm{camp}=260\,\mathrm{d}$, 
the average time span between observations $t_\mathrm{cyc}=2\,\mathrm{d}$,
the exposure time per night $t_\mathrm{exp}=0.5\,\mathrm{h}$, 
the photon detection rate $\epsilon = 43\,\mathrm{phot/s}$, 
the median seeing $\theta_\mathrm{see}=1.5\,\mathrm{arcsec}$, 
the area of the image $\Omega_\mathrm{ccd}=70\,\mathrm{arcmin}^2$, the 
inaccuracy of the PSF matching $\eta_\mathrm{psf}^2=0$, and the factor in 
the noise due to sky photons $\eta_\mathrm{sky}^2=2.6$.
\\
This calculation can only serve as a crude estimate of the event rate.
In our subsequent papers we will present calculations more appropriate for 
our data set.

\section{The WeCAPP Project}

\subsection{Telescopes and Instruments}

The Wendelstein 0.8\,m telescope has a focal length $f$ of 9.9\,m, which 
results in an aperture ratio $f/D=12.4$. Starting in September 1997 we used 
a TEK CCD with $1024\times1024$ pixels of $24\,\mu\mathrm{m}$ corresponding 
to 0.5\,arcsec on the sky. With this CCD chip we were able to cover
$8.3\times8.3\,\mathrm{arcmin}^2$ of the bulge of M\,31. 
To increase the time sampling of our observations we started to 
use the Calar Alto 1.23\,m telescope ($f=9.8\,\mathrm{m}$, $f/D=8.0$) in 1999.
The observations were partly carried out in service mode. Six different 
CCD chips were used. Three of these CCDs cover a field of 
$17.2\times17.2\,\mathrm{arcmin}^2$ 
and were used to survey the whole bulge for 
lensing events. A detailed overview of the properties of each CCD 
camera used for WeCAPP is given in Table \ref{ccd}.
\begin{table*}
\centering
\begin{tabular}{llcccccc}
\hline
\hline
Site & Campaign & CCD  & Size & [$\mathrm{arcsec}/\mathrm{px}$] & Field [$\mathrm{arcmin}^2$]
& \# of R frames & \# of I frames \\
\hline
We & 1997/1998 & TEK\#1      & 1K $\times$ 1K & 0.49 & $8.3   \times   8.3$ &  276 & 123 \\
We & 1998/1999 & TEK\#1      & 1K $\times$ 1K & 0.49 & $8.3   \times   8.3$ &  454 & 210 \\
We & 1999/2000 & TEK\#1      & 1K $\times$ 1K & 0.49 & $8.3   \times   8.3$ &  835 & 358 \\
CA & 1999/2000 & TEK7c\_12   & 1K $\times$ 1K & 0.50 & $8.6   \times   8.6$ &  266 & 136 \\ 
CA & 1999/2000 & TEK13c\_15  & 1K $\times$ 1K & 0.50 & $8.6   \times   8.6$ &   62 &  33 \\ 
CA & 1999/2000 & SITe2b\_11  & 2K $\times$ 2K & 0.50 & $17.2  \times  17.2$ &   14 &   7 \\
CA & 1999/2000 & SITe2b\_17  & 2K $\times$ 2K & 0.50 & $17.2  \times  17.2$ &  448 & 249 \\
CA & 1999/2000 & SITe18b\_11 & 2K $\times$ 2K & 0.50 & $17.2  \times  17.2$ &  165 &  91 \\
CA & 1999/2000 & LOR11i\_12  & 2K $\times$ 2K & 0.31 & $10.75 \times 10.75$ &  219 &  92 \\
\hline\\
\end{tabular}
\caption{ Properties of all CCD cameras used during WeCAPP at 
Wendelstein (We) and Calar Alto (CA) Observatories, respectively. 
All CCDs have a pixel size of $24\,\mu\mathrm{m}$, except the Loral
 which has $15\,\mu\mathrm{m}$ 
pixels.}
\label{ccd}
\end{table*}
\\
Most of the sources for possible lensing events in the bulge of M\,31
are luminous red stars i.e. giants and supergiants. Consequently the
filters used in our project should be sensitive especially to these
kind of stars. We chose therefore R and I filters for our survey.
At Wendelstein we used the R2 
($\lambda\simeq650\,\mathrm{nm}$,
$\Delta\lambda\simeq150\,\mathrm{nm}$)
and Johnson I 
($\lambda\simeq850\,\mathrm{nm}$,
$\Delta\lambda\simeq150\,\mathrm{nm}$)
wavebands. 
To be as consistent as possible 
with the data obtained at Wendelstein, the Calar Alto observations were 
carried out with the equivalents filters, R2 
($\lambda\simeq640\,\mathrm{nm}$,
$\Delta\lambda\simeq150\,\mathrm{nm}$)
and Johnson I 
($\lambda\simeq850\,\mathrm{nm}$,
$\Delta\lambda\simeq150\,\mathrm{nm}$).
Since June 2000 we are using the newly installed filters Johnson R 
($\lambda\simeq640\,\mathrm{nm}$, 
$\Delta\lambda\simeq160\,\mathrm{nm}$)
and Johnson I 
($\lambda\simeq850\,\mathrm{nm}$, 
$\Delta\lambda\simeq150\,\mathrm{nm}$)
at Calar Alto.
\\
Despite of the combination of different telescopes, CCDs, and slightly
different filter systems we observed no systematic effects in the light curves
depending on these parameters.

\subsection{Observing Strategy}

\begin{figure}
\centering
\includegraphics[width=0.43\textwidth]{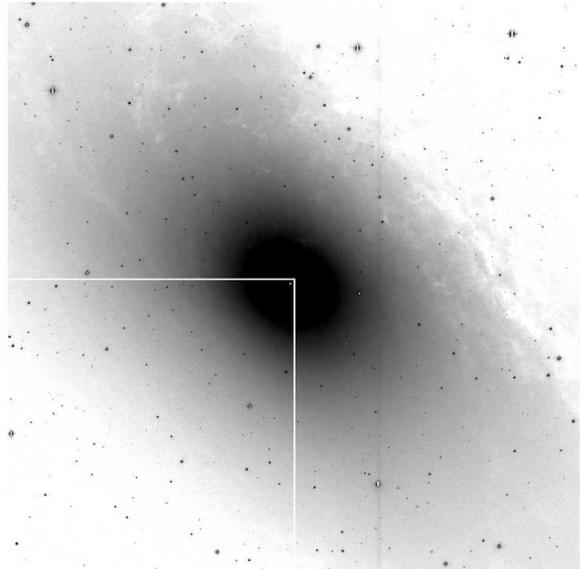}
\caption{M\,31 frame observed on 2000 June 26th with the Calar Alto
1.23 m telescope: $17.2 \times 17.2\;\mathrm{arcmin}^2$
($3.85 \times 3.85\;\mathrm{kpc}^2$).
A quarter of this field, marked by the solid lines, was observed during
the first three campaigns from September 1997 until March 2000.
\vspace*{0.5cm}
}
\label{m31}
\end{figure}

To follow the suggestion of \citet{1996AJ....112.2872T} and 
\citet{1996ApJ...473..230H} we chose the field with the maximal lensing 
probability, pointing to the far side of the M\,31 disk. The main fraction 
of the field is covered by the bulge of M\,31 with the nucleus of M\,31 
located at one corner of the field (Fig.~\ref{m31}).
As gravitational lensing is achromatic, the amplification of the source is 
the same in different wavebands. However, as shown in several papers
\citep[e.g.][]{1995LSS..1..326V,1995ApJ...449...42W,2000MNRAS.316...97H}
blending on the one hand and differential 
amplification of an extended source on the other hand can lead to a 
chromatic, but still symmetric, lensing light curve. Under certain
circumstances chromatic light curves permit to constrain the physical 
properties of the source-lens system \citep[e.g.][]{1996ApJ...464..212G,
2001MNRAS.320...41H}. 
Variable stars will generally change colour in a different way.
Our observation cycle therefore comprises 5 images in the R band and
3 images in the I band lasting about 45 min including readout
time. Stacking these images with an average exposure time of 150 sec 
in R and 200 sec in I results in a magnitude limit between 
(20.8 -- 22.1) mag in R and (19.1 -- 20.4) mag in I for a point source 
on the background of M\,31 and a signal-to-noise ratio 
$(S/N)=10$ in over 95\,\% of the frame.
The background of M\,31 typically has a surface brightness between
(18.7 -- 21.2) $\mathrm{mag}/\mathrm{arcsec}^2$ in R and (16.8 -- 19.3) 
$\mathrm{mag}/\mathrm{arcsec}^2$ in I. The cycles were repeated as often as 
possible during one night, usually at least twice. As we had to 
avoid saturation of stars in the observed field we made exposure 
times dependent of the actual seeing, whereas exposure times
in the I bands where generally longer.

\subsection{The Data}

\begin{figure*}
\centering
\includegraphics[width=0.3\textwidth]{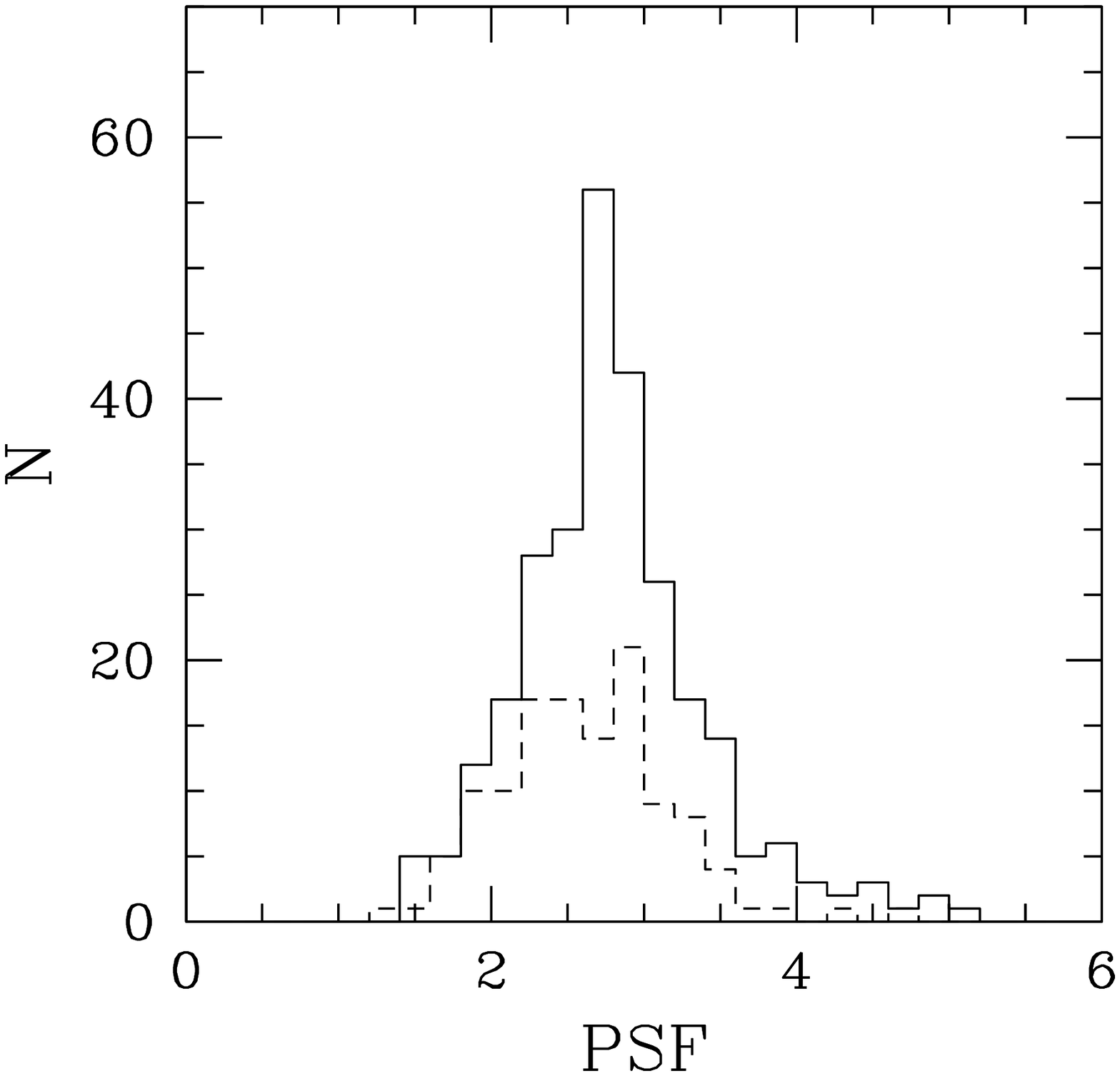}
\includegraphics[width=0.3\textwidth]{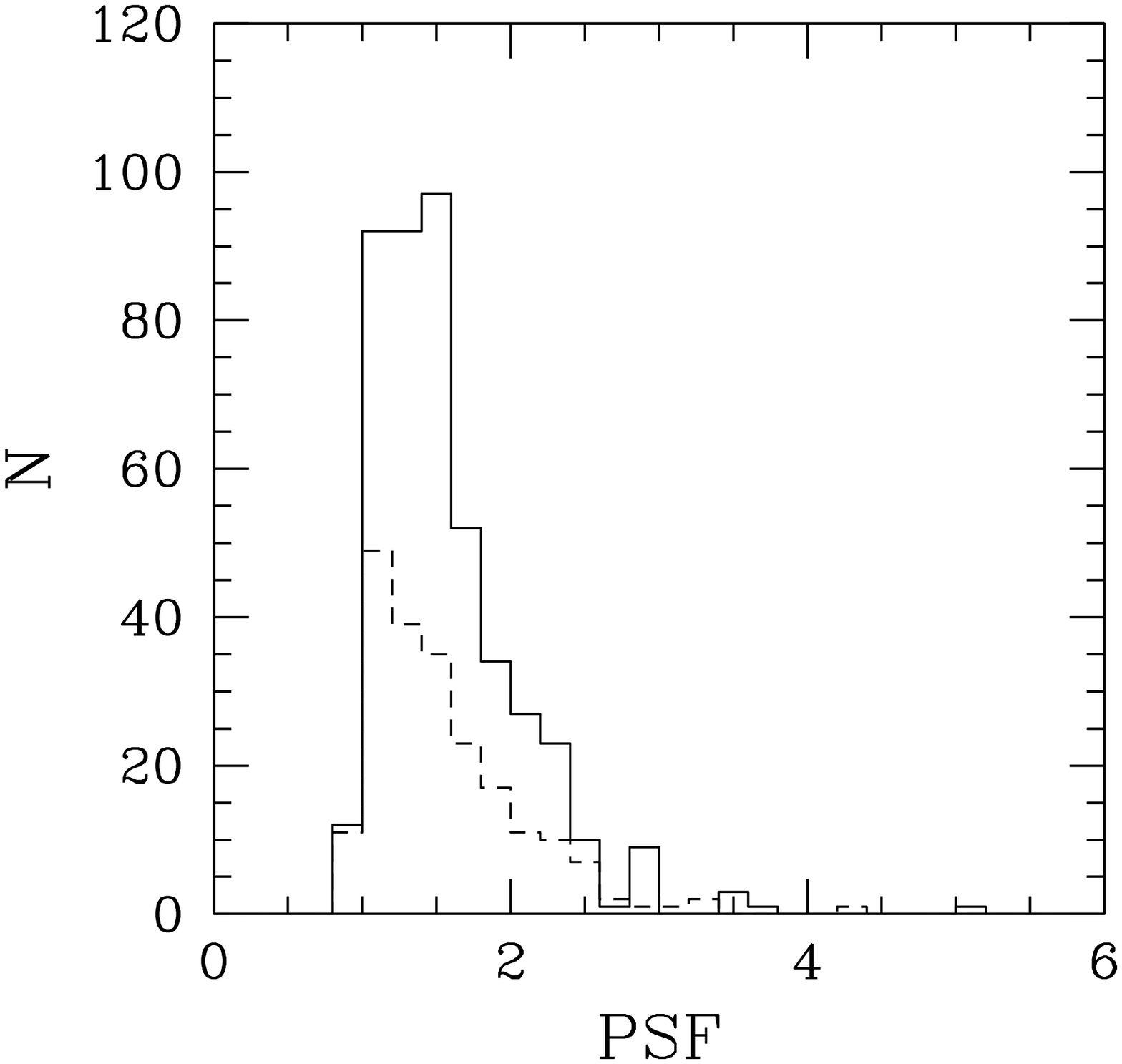}
\caption{Histograms of the Full Width Half Maximum 
(FWHM) of the point spread function (PSF) of the frames taken at Wendelstein 
Observatory during the 1997/1998 campaign (left panel) and the 1998/1999 
campaign (right panel). Frames in the R band are marked by a solid line, 
frames in the I band by a dashed line. The lower limit of the PSF is 
restricted by a pixel size of 0.5 arcsec.}
\label{ws_97_99}
\end{figure*}

\begin{figure*}
\centering
\includegraphics[width=0.3\textwidth]{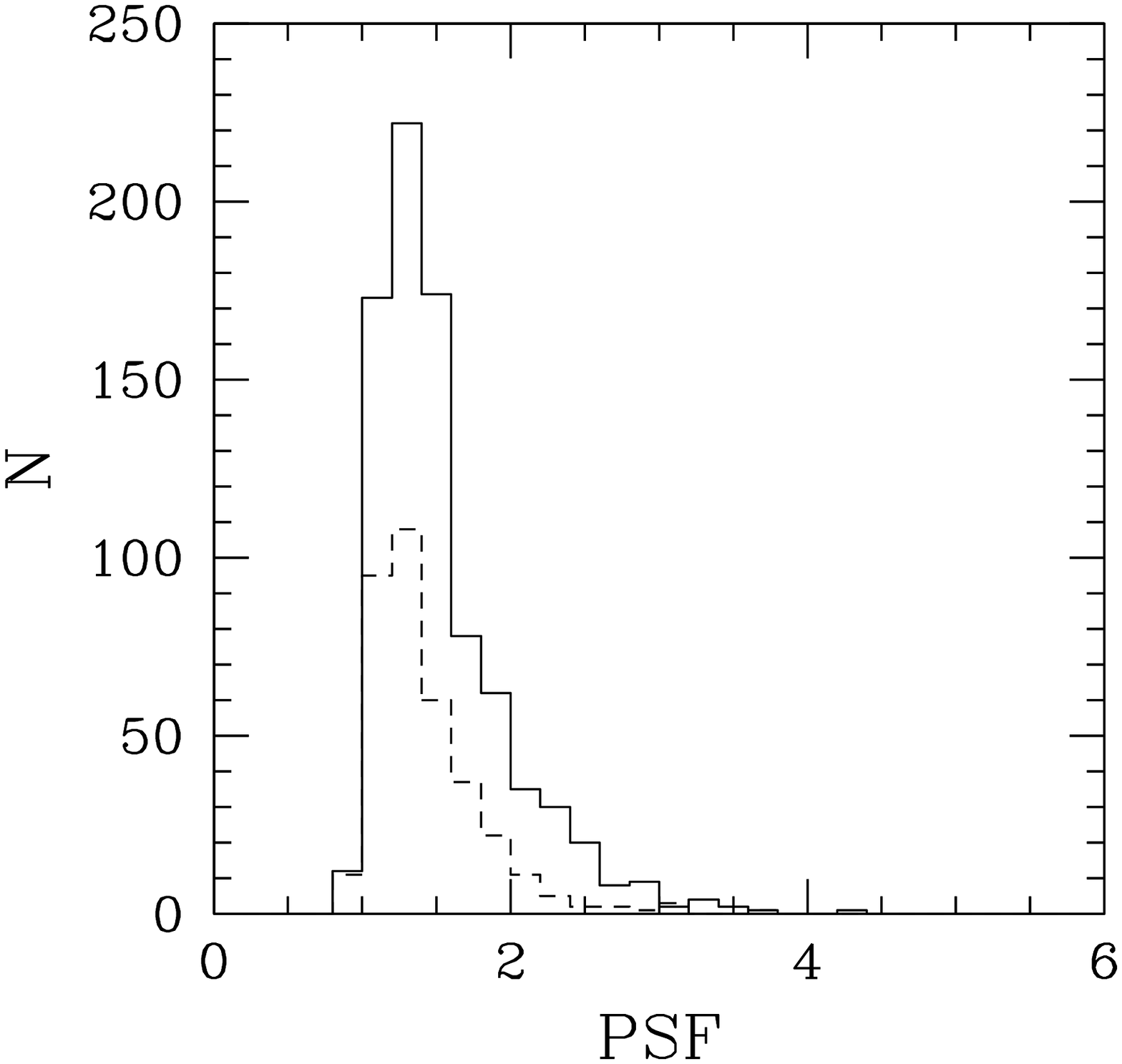}
\includegraphics[width=0.3\textwidth]{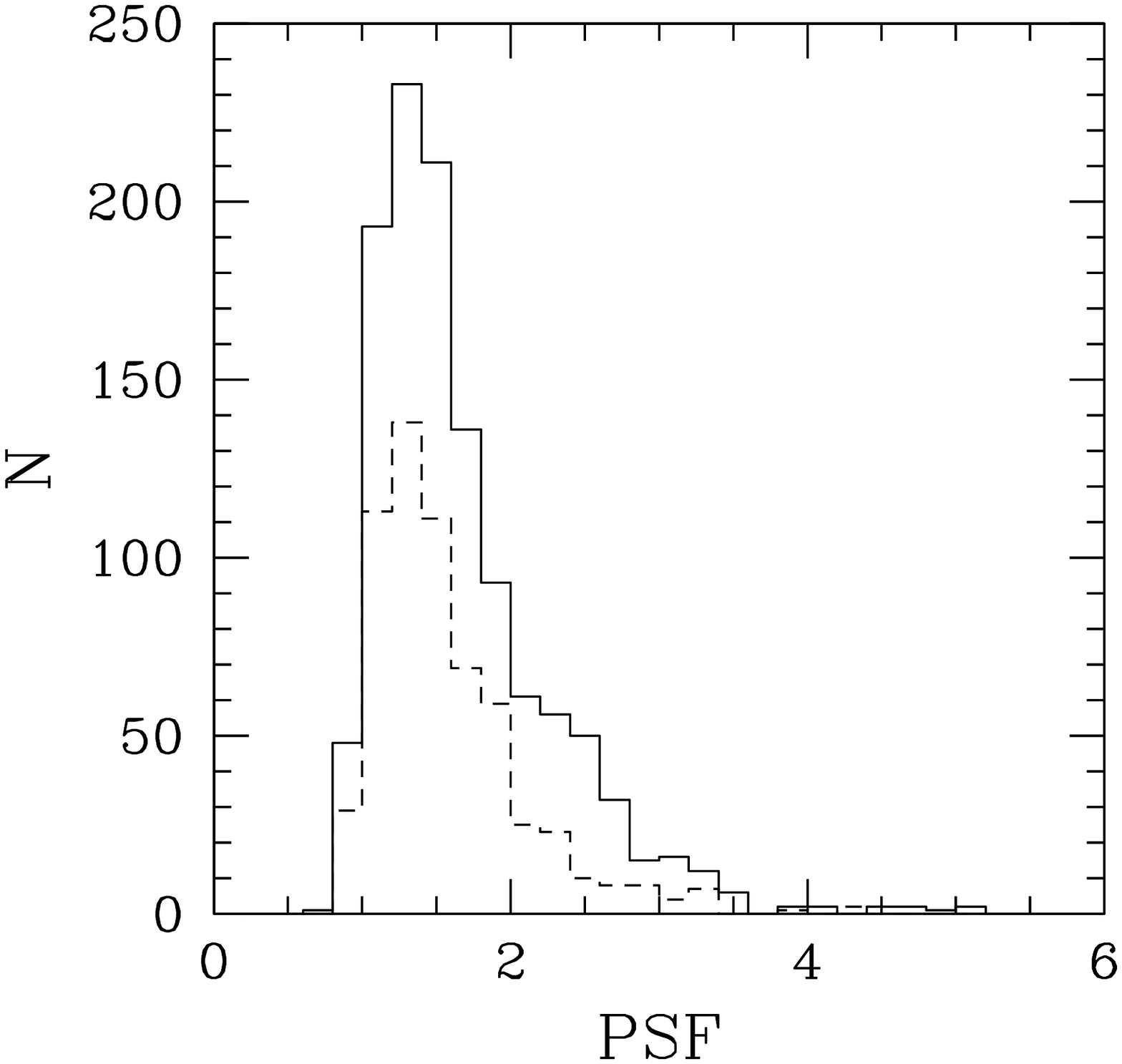}
\caption{Histograms of the FWHM of the frames taken 
during the 1999/2000 campaign at Wendelstein (left panel) and Calar Alto 
Observatory (right panel). Frames in the R band are marked by a solid 
line, frames in the I band by a dashed line. Note that the pixel sizes 
of the CCD cameras used correspond to 0.5 arcsec on the sky.}
\label{cam_99_00}
\end{figure*}

We began our observations at Wendelstein with a test period
in September 1997, observing on 35 nights until March 1998. The second
observational period lasted from 1998 October 22nd until 1999 March 24th .
During the first Calar Alto campaign we received two hours of service 
observations on 87 of 196 allocated nights (1999 June 27th - 2000 March 3rd). 
From  November 1st until November 14th we were able to observe during the 
whole night. In parallel we continued our observations at Wendelstein 
on 221 nights, of which 65 were clear. In this way we achieved an overall 
time coverage of 132 nights (52.6\,\%). During the three years of WeCAPP 
we collected at Wendelstein Observatory a total of 1565 images in the 
R band and 691 images in the I band. The observations for over more 
than one year at Calar Alto resulted in 1174 frames in the R and 608 
frames in the I band. 
\\
During the 1997/1998 test campaign conditions at the Wendelstein
telescope were improved significantly. A new\-ly installed air conditioning
system reduced dome seeing to a low level. Further improvements like 
fans just above the main mirror finally lead to a leap in the image 
quality obtained with the telescope. Figure~\ref{ws_97_99} which presents 
the PSF statistics of Wendelstein images from the 1997/1998 and 1998/1999 
campaigns respectively illustrates this fact.
In general Wendelstein shows a marginally better PSF distribution than 
Calar Alto (see Figs.~\ref{ws_97_99} and \ref{cam_99_00}). Table \ref{med}
shows the PSF median values for the images taken during WeCAPP at both sites.
\\
\begin{table*}
\centering
\begin{tabular}{ccccc}
\hline
\hline
Filter & We 1997/1998 & We 1998/1999 & We 1999/2000 & CA 1999/2000 \\
\hline
R &  2.76 & 1.45 & 1.40 &  1.49 \\
I &  2.59 & 1.44 & 1.32 &  1.44 \\
\hline\\
\end{tabular}
\caption{Median values of the FWHM of the PSF, given in arcsec, 
for the images taken during WeCAPP in the R and I band at Wendelstein 
and Calar Alto Observatories.}
\label{med}
\end{table*}
\begin{figure*}
\centering
\includegraphics[width=0.7\textwidth]{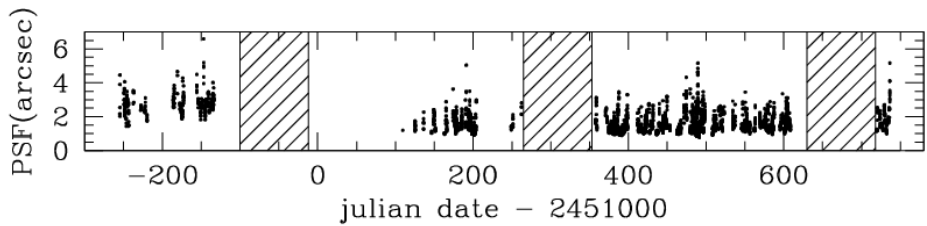}
\caption{Illustration of PSF vs. time coverage during three 
years of WeCAPP. Shaded regions mark the periods of time when M\,31 was not 
observable.}
\label{sam}
\end{figure*}
Figure~\ref{sam} shows the time sampling we reached so far with WeCAPP. 
Because of time loss during the upgrades of the telescope, time
coverage of the 1997/1998 campaign is only fragmentary. About the same 
applies to the following campaign, this time due to a
camera shutdown and another time consuming project.
Finally time coverage of the first joint campaign of Wendelstein and
Calar Alto is good, last but not least due to the often opposite
weather situation in Spain and Germany.

\section{Data Reduction}

\subsection{Reduction Pipeline}

In the last three years we developed an image reduction pipeline
being able to cope with a massive imaging campaign. This reduction 
pipeline is described in detail in \citet{2001AA..00..00} and 
combines all reduction steps from de-biasing of the images until the final 
measurements of the light curves in one software package, including full
error propagation from the first reduction step to the last:
\begin{enumerate}
\item standard CCD reduction including de-biasing, flat-fielding and
filtering of cosmics 
\item position alignment using a 16 parameter interpolation
\item stacking of frames
\item photometric alignment
\item PSF matching using OIS (Optimal Image Subtrac\-tion), a method
proposed by \citet{1998ApJ...503..325A}
\item generation of difference images
\item detection of variable sources
\item photometry of the variable sources
\end{enumerate}

\subsection{Standard CCD Reduction}

Pre-reduction of the raw frames is performed in a standard way:
After de-biasing of the frames, saturated and bad pixels are marked.
We use the 3 $\sigma$ clipped median of a stack of at least 5 twilight 
flatfields for the flat fielding procedure.
\\
Cosmics are effectively detected by fitting a Gaussian PSF to all local maxima 
in the frame. All PSFs with a FWHM lower than 0.7 arcsec and an amplitude of 8 
times the background noise are removed. This provides a very reliable 
identification and cleaning of cosmics.

\subsection{Position Alignment}

After determining the coordinates of the reference objects by a PSF fit, 
we calculate a linear coordinate transformation to project an image 
onto the position of the reference frame.
\\
The flux interpolation for non-integer coordinate shifts 
is calculated from a 16-parameter, $3^\mathrm{rd}$-order polynomial
interpolation using 16 pixel base points.

\subsection{Stacking Frames}

To avoid saturation of Galactic foreground stars and the nucleus of M\,31 
in our field, exposure times were limited to a few hundred seconds.
Therefore one has to add several frames taken in one cycle to obtain an 
acceptable signal-to-noise ratio $(S/N)$. Usually we 
stacked 5 frames in the R band and 3 frames in the I  according to 
two criteria, comparable PSF and comparable sky. Frames with very high 
background levels or very broad PSFs were not added if they reduced the 
detectability of faint variable sources.
Consequently the number of images to be stacked was not fixed, coaddition of 
frames was performed in a way to get a maximum $(S/N)$ 
ratio for faint point sources in the stacked frame. 
\\
Before OIS is finally carried out we align all frames photometrically.
This ensures that all light curves are 
photometrically calibrated to a standard flux.

\subsection{PSF Matching}

\begin{figure*}
\centering
\includegraphics[width=0.4\textwidth]{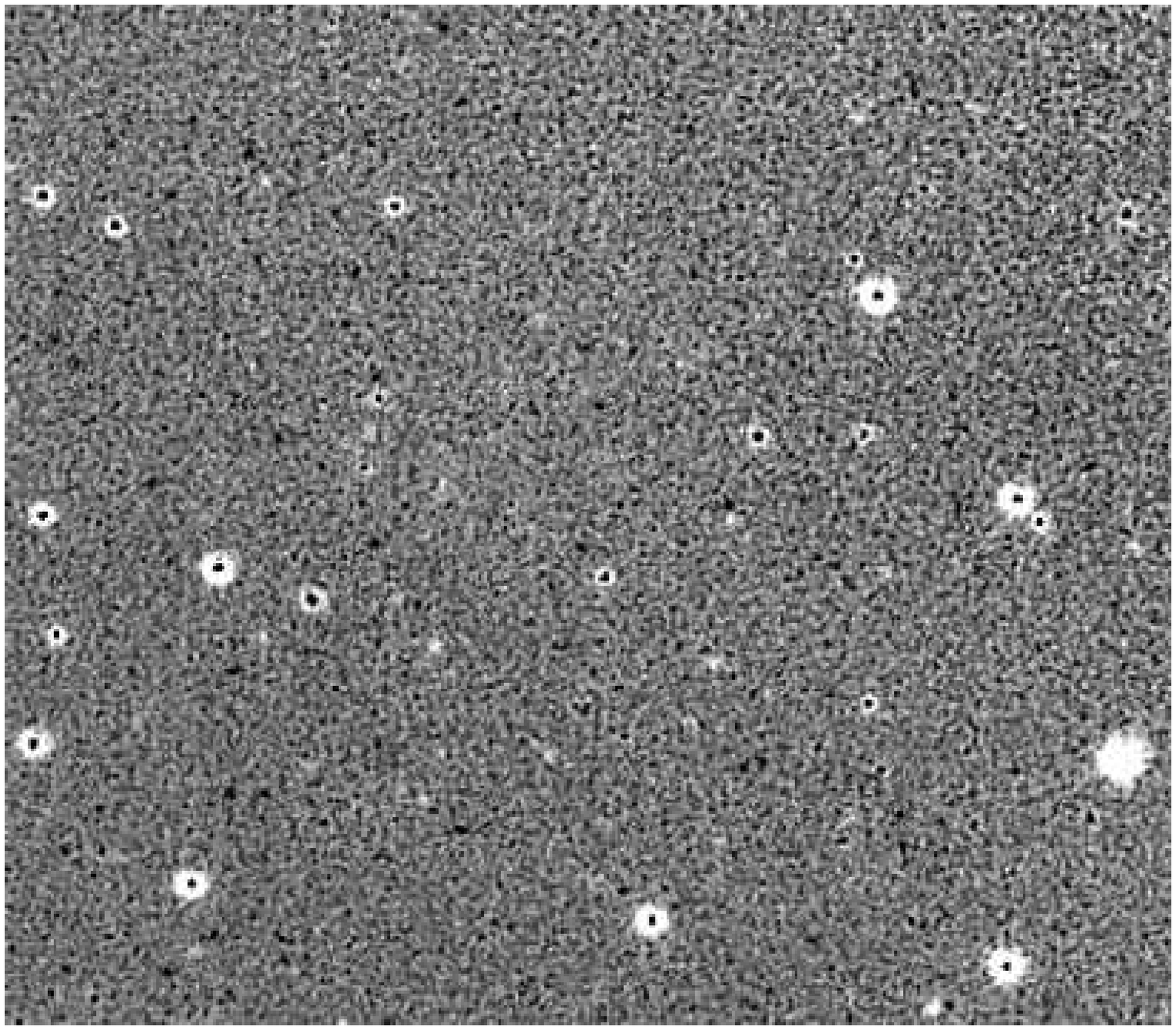}
\includegraphics[width=0.4\textwidth]{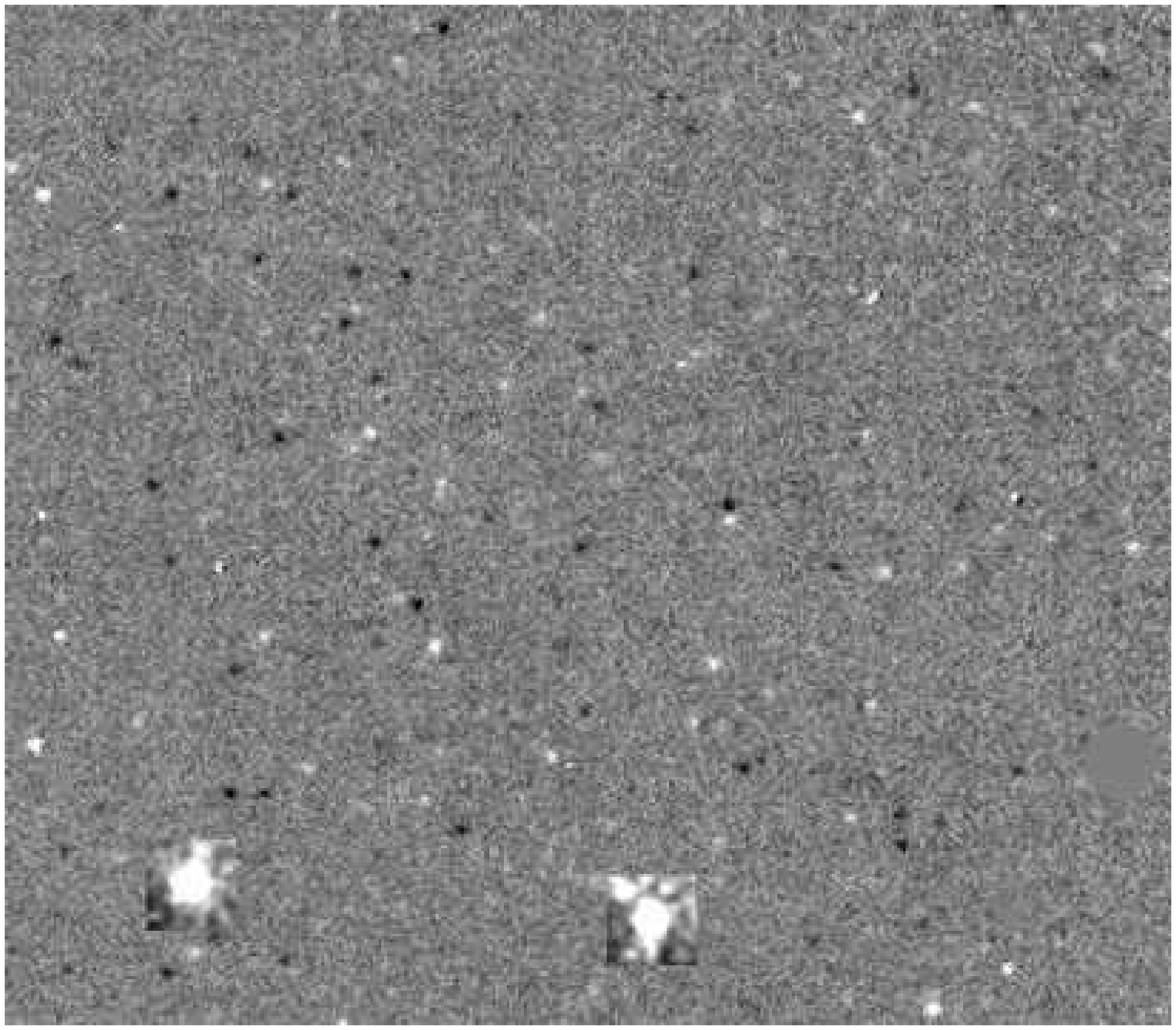}
\caption{Difference images of a part of the M\,31 bulge 
($3.25\times2.82\,\mathrm{arcmin}^2$), 
centered at 6.6 arcmin distance from the nucleus.
Left panel: difference frame without OIS. Because of the relatively
large residuals from stars no identification of faint variable stars 
is possible. Right panel: difference frame with OIS. Bright and dark spots
are variable sources. The two boxes represent stars not subtracted in order
to provide information on the PSF.}
\label{diff}
\end{figure*}

In order to extract light curves of variable sources from the data
we use a method called Difference Image Analysis (DIA), proposed by 
\citet{1990PASP..102.1113C} and first implemented by 
\citet{1996AJ....112.2872T} in a lensing study.
\\
The idea of DIA is to subtract two positionally and photometrically
aligned frames which are identical except for variable sources.
The resulting difference image should than be a flat noise frame,
in which only the variable point sources are visible.
\\
The crucial point of this technique apart from position
registration is the requirement of a perfect matching of
the point spread functions (PSFs) between the two frames.
\\
The PSF of the reference frame $r$ is convolved with a kernel $k$ 
to match the broader PSF of an image $i$,
\begin{equation}
i(x,y) \approx r(u,v) \otimes k(u,v) + bg(x,y) \equiv \tilde{r}(x,y) 
\end{equation}
where $bg$ accounts for the background level and $\tilde{r}$ is the 
transformed reference frame.
\\
In order to obtain an optimal kernel $k$ we implemented OIS as proposed by 
\citet{1998ApJ...503..325A}. This least-squares fitting method determines $k$ 
by decomposing it into a set of basis functions.
We use a combination of three Gaussians with different 
widths $\sigma$ multiplied with polynomials up to $6^\mathrm{th}$ order. 
This leads to the following $49$ parameter decomposition 
of $k(u,v)$:
\begin{equation}
\begin{array}{rl}
  \sigma_1=1: & e^{-\frac{u^2+v^2}{2\sigma_1^2}} \;
                (a_1 + \dots + a_{22} u^6 + \dots + a_{28} v^6) \\ 
  \sigma_2=3: & e^{-\frac{u^2+v^2}{2\sigma_2^2}} \;
                (a_{29}+...+a_{39} u^4+...+a_{43} v^4) \\
  \sigma_3=9: & e^{-\frac{u^2+v^2}{2\sigma_3^2}} \;
                (a_{44} + \dots + a_{47} u^2 + a_{48} uv + a_{49} v^2)
\end{array}
\end{equation}
Additionally 3 parameters are used to fit the background
\begin{equation}
bg(x,y) = a_{50} + a_{51}\,x + a_{52}\,y \; .
\end{equation}
To cope with the problem of a PSF varying over the area
of the CCD we divide the images in sub-areas of 141 x 141 pixels each.
In all sub-areas a locally valid convolution kernel is
calculated.
As we have chosen the kernel to have 21 x 21 pixels 
we therefore effectivly use 161 x 161 to derive $k(u,v)$.
\\
Differential refraction causes a star's PSF to depend 
on its colour \citep[chap. 4.4]{1996AJ....112.2872T}. 
However these second order 
effects are negligible for our data set and do not lead to residuals in the 
difference images.
\\
As we are performing DIA we have to choose a 
reference frame $r$ which will be subtracted from all other coadded 
frames $i$ and which determines the baseline of the light curve.
OIS shows best results for a small PSF and a high $(S/N)$  
reference frame. Therefore, the best stacked images were coadded once more. 
Our actual R band reference frame comprises 20 images
taken at 2 different nights resulting in a total exposure time of 2400
seconds and a PSF of a FWHM of 1.05 arsec (= 2 pixels for all CCDs except
one at Wendelstein and Calar Alto).  For the I band we coadded 
18 frames (i.e. 6 stacked frames taken at 3 different nights) 
which results in a total exposure time of 3160 seconds and a FWHM 
of 1.15 arcsec.
As we are continuing collecting data the process of constructing 
the ultimate reference frame has not finished yet. Each night of 
high quality data collected at one of the two sites will improve the
reference frame further. Figure~\ref{diff} shows a typical difference 
image obtained by using our implementation of OIS.

\subsection{Source Detection}

To detect sources in the difference images 
we fit a rotated Moffat function \citep{1969A&A.....3..455M}
to all local maxima in the binned frame.
Real sources are filtered by rejecting sources with an amplitude less 
than 5 times the background noise.

\subsection{Photometry of the variable sources}

Photometry of the detected sources is performed by a profile 
fitting technique. To obtain information about the PSF of any 
particular frame we apply a Moffat fit \citep{1969A&A.....3..455M}
to several reference stars in the CCD field. 
Having determined the shape of the PSF, we perform Moffat fits on the positions
of the variable sources as returned from the detection algorithm. 
In these final fits the amplitude is the only free parameter. 
To determine the flux of the source we finally integrate the count 
rates over the area of the (now fully known) analytical function of 
the PSF. This minimizes the contamination from neighbouring sources.

\subsection{Calibration of the Light Curves}

As the coadded images are normalized to the reference frame it is not 
necessary to calibrate each image separately. Only the reference frame 
is calibrated once.
\\
To calculate magnitudes $m_\mathrm{R'}$ in our $\mathrm{R'}$ band, which 
corresponds to 
R2, we determined the instrumental zeropoint $ZP_\mathrm{R'}$ and
the extinction coefficient $\kappa$: 
\begin{equation}
  m_\mathrm{R'} = -2.5\log[ADU/t] - \kappa \cdot AM + ZP_\mathrm{R'}
  \label{calib}
\end{equation}
where $t$ is the exposure time and $AM$ is the airmass. 
\\
Aperture photometry with 7 different Landolt standard stars 
\citep{1992AJ....104..340L} observed
at different airmasses was performed for a photometric night
at Calar Alto Observatory. With these
stars the extinction $\kappa$ for the night was calculated to
$\kappa_\mathrm{R'}=0.073\pm0.005$. 
To determine the zeropoint for the $\mathrm{R'}$ band 
we used an A0V-star, \object{Feige 16}, with the
colours $(B-V)=-0.012$, $(U-B)=0.009$, $(V-R)=-0.003$, 
$(R-I)=0.002$,
and a visual magnitude of $V=12.406$ mag. The zeropoint was
determined according to Eq. \ref{calib} to $ZP_\mathrm{R'}=23.05\pm0.02\ 
\mathrm{mag}$ and used to calculate the magnitudes for the reference frame. 
This zeropoint is not valid for Wendelstein.
\\
In the following, we only give fluxes for the sources in our filter system, 
because the intrinsic magnitudes and colours of our unresolved 
sources cannot be determined with sufficient accuracy.
\\
We show the light curves in flux differences according to
\begin{equation}
  \Delta f_\mathrm{R'}= f_\mathrm{R', Vega}
      \frac{\Delta ADU_\mathrm{R'}}{t} 10^{0.4\,\kappa \ \mathrm{AM}} 
      10^{-0.4\,ZP_\mathrm{R'}}
\end{equation}
with $f_\mathrm{R',Vega} = 3124\,\mathrm{Jy}$ from an integration over 
the CCD-filter system. 
\\
The same transformations were done for our $\mathrm{I'}$ band, 
corresponding to Johnson I (Calar Alto), 
with $\kappa_\mathrm{I'}=0.025\pm0.005$, 
$ZP_\mathrm{I'}=21.82\pm0.03\,\mathrm{mag}$ and 
$f_\mathrm{I',Vega} = 2299\,\mathrm{Jy}$.
\\
The colour terms between all filter sets we used in our observation are 
negligible. The transformation to the standard Kron-Cousins filter
system is: $R=R'+ZP_\mathrm{R'}+0.06(R-I)$, 
$I=I'+ZP_\mathrm{I'}+0.38 (R-I)$.

\section{Results}

\begin{figure}
\centering
\includegraphics[width=0.40\textwidth]{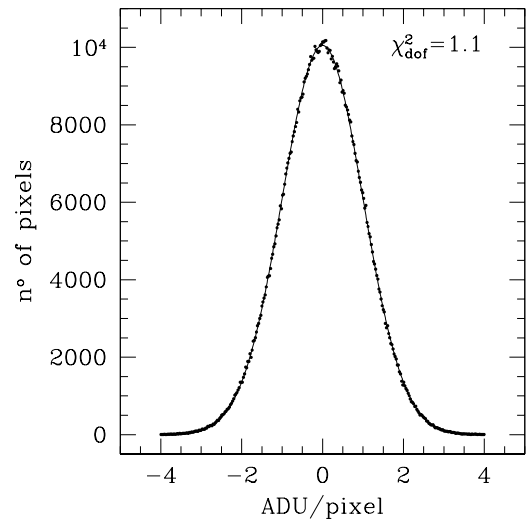}
\caption{Histogram of the pixel values of a simulated 
difference image $d_{xy}=i_{xy}-r_{uv}\otimes k_{uv}$ ($r$ is the
reference frame, $i$ is a second frame and $k$ is the convolution kernel).
The pixel values are divided by the expected rms errors, as derived
from error propagation. The solid curve shows a Gaussian with $\sigma=1$.
We calculated the reduced chi-square $\chi^2_\mathrm{dof}$ of 
19 different simulated images in the range between -3 and 3. 
The median $\bar{\chi}^2_\mathrm{dof}$ is 1.1, which means that expected 
and measured errors match almost perfectly and
that the residuals of the OIS are at the photon noise level.}
\label{photon_noise} 
\end{figure}

\begin{figure}
\centering
\includegraphics[width=0.48\textwidth]{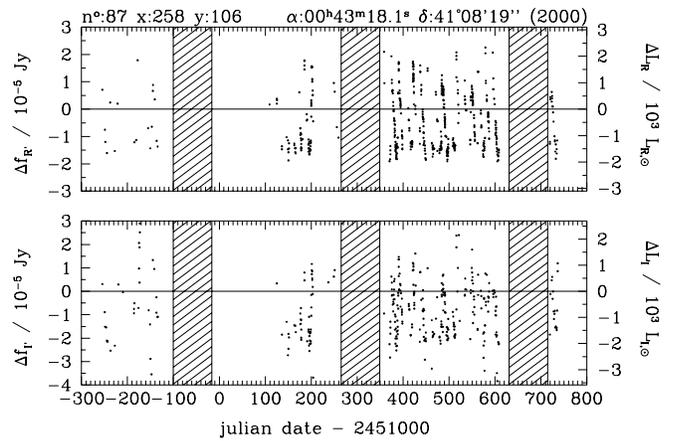}
\caption{Light curve of a $\delta$-Cephei variable, 
upper panel: $\mathrm{R'}$ band, lower panel: $\mathrm{I'}$ band.}
\label{lc_01} 
\end{figure}

\begin{figure*}
\centering
\includegraphics[width=0.32\textwidth]{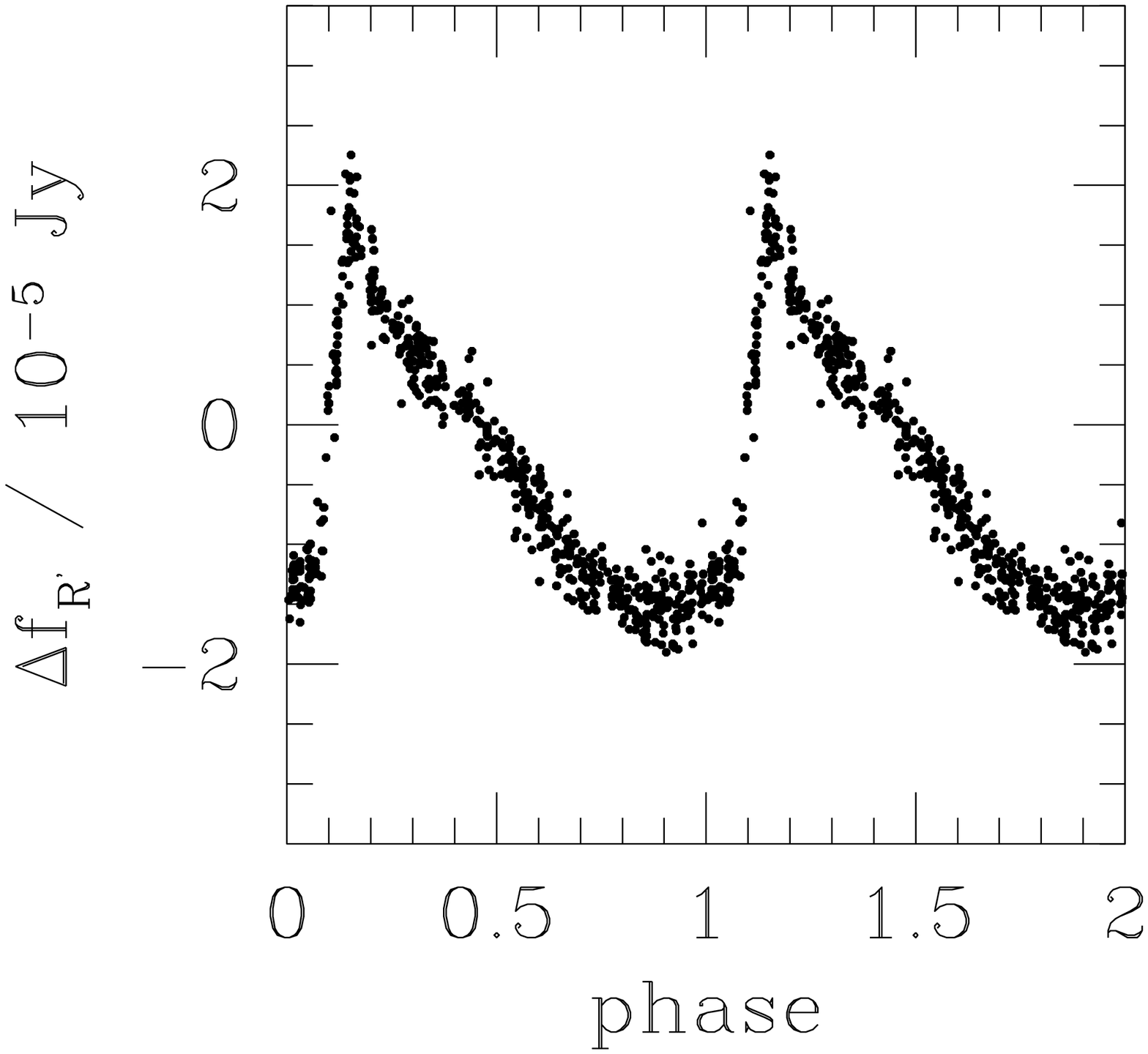}
\includegraphics[width=0.32\textwidth]{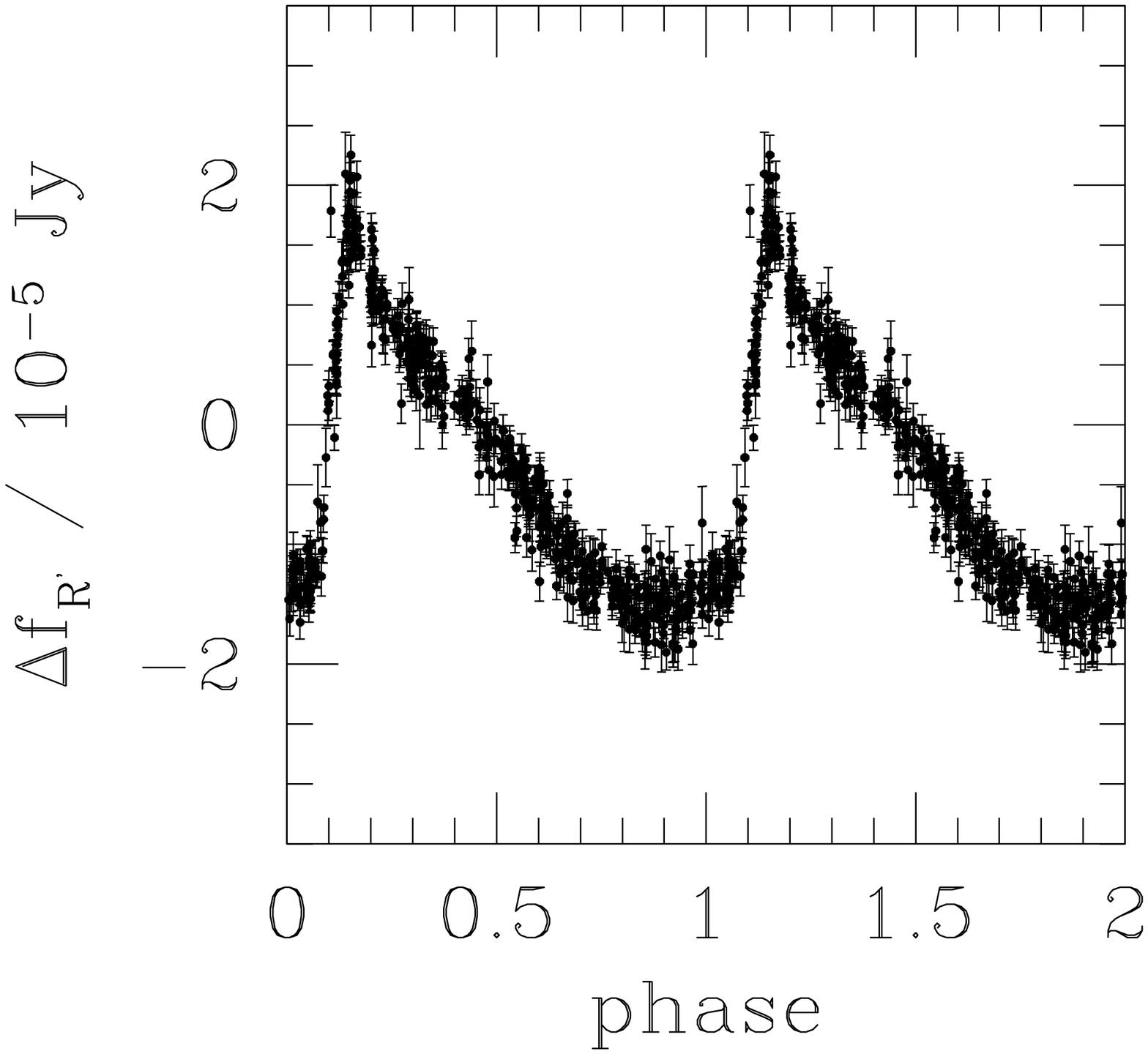} 
\includegraphics[width=0.32\textwidth]{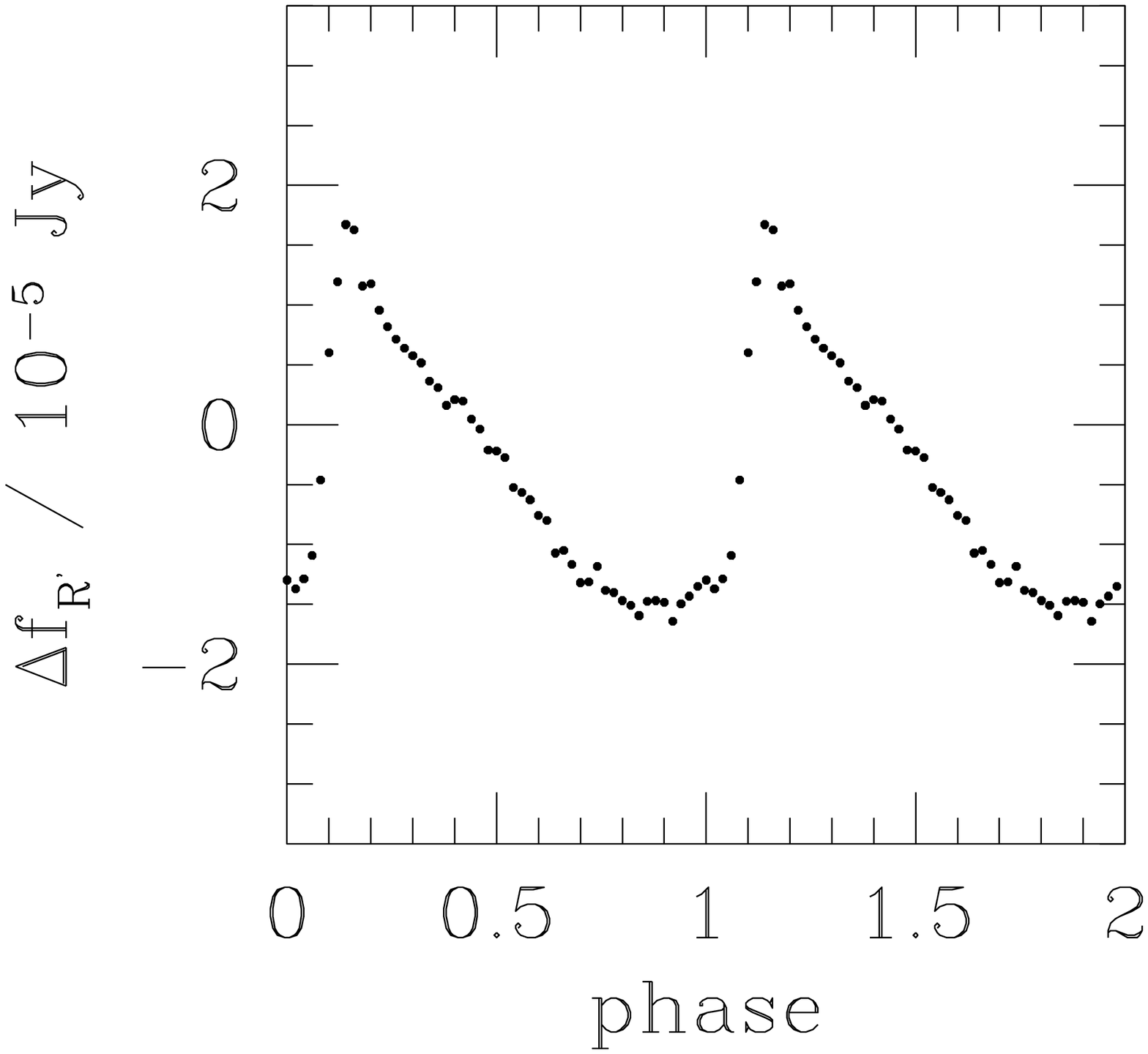} 
\caption{Light curve of the $\delta$-Cephei 
star of Fig.~\ref{lc_01} in the $\mathrm{R'}$ band, convolved with its 
period of 
$P=15.76\pm0.01$ d. Plotted without (left panel) and with (centre panel) 
$1\sigma$ error bars, which represent fully propagated errors through 
all reduction steps. Right panel: Binned $\mathrm{R'}$ 
light curve of this star.}
\label{convol1} 
\end{figure*}

\begin{figure}
\centering
\includegraphics[width=0.48\textwidth]{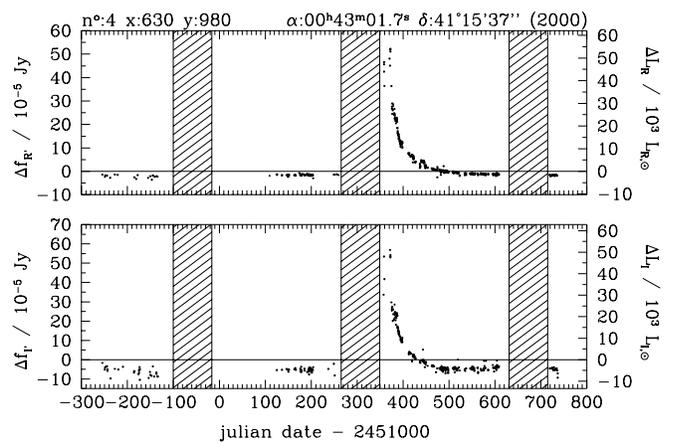}
\caption{Light curve of a nova, representing the brightest 
variable source detected in our M\,31-field. This nova was previously 
published by \citet{1999IAUC.7218....2M}. Upper panel: $\mathrm{R'}$-Band, 
lower panel: $\mathrm{I'}$-Band.}
\label{lc_02} 
\end{figure}

\begin{figure}
\centering
\includegraphics[width=0.48\textwidth]{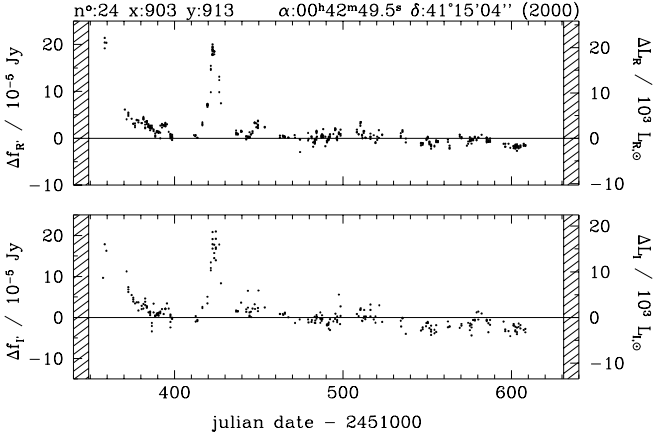}
\caption{Light curve of an eruptive variable, which could 
be mistaken as a microlensing event, if the time coverage were 
insufficient. Upper panel: $\mathrm{R'}$ band, lower panel: $\mathrm{I'}$ 
band.}
\label{lc_09} 
\end{figure}

\begin{figure}
\centering
\includegraphics[width=0.48\textwidth]{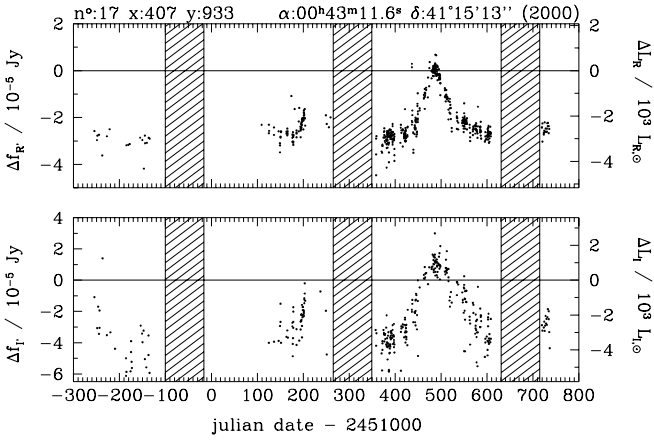}
\caption{Light curve of a longperiodic variable. 
Upper panel: $\mathrm{R'}$ band, lower panel: $\mathrm{I'}$ band. Note, that 
insufficient time coverage could result
in a false identification of this variable as a microlensing event.}
\label{lc_03} 
\end{figure}

\begin{figure}
\centering
\includegraphics[width=0.48\textwidth]{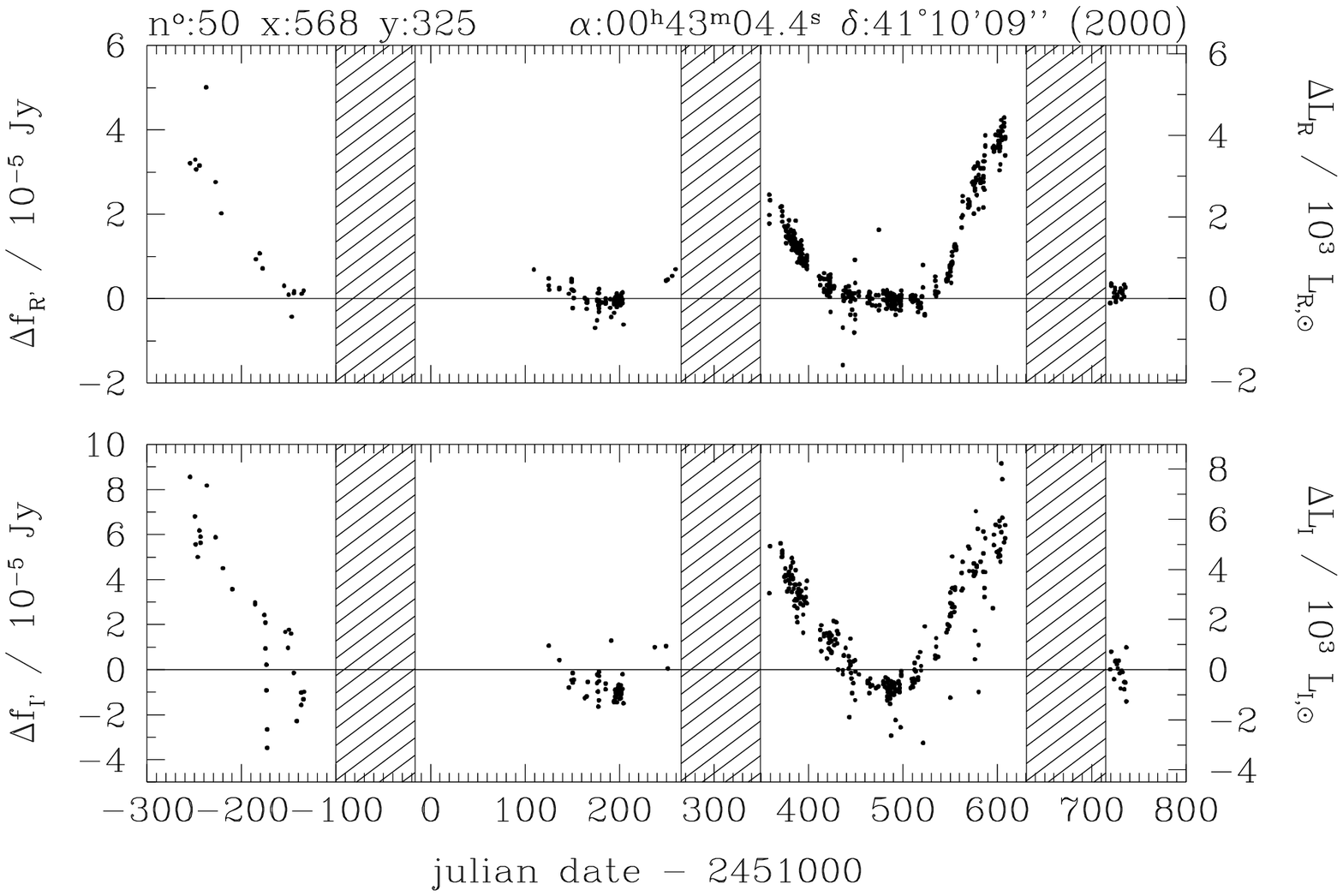}
\caption{Light curve of a longperiodic variable. 
Upper panel: $\mathrm{R'}$ band, lower panel: $\mathrm{I'}$ band.}
\label{lc_04} 
\end{figure}

\begin{figure}
\centering
\includegraphics[width=0.48\textwidth]{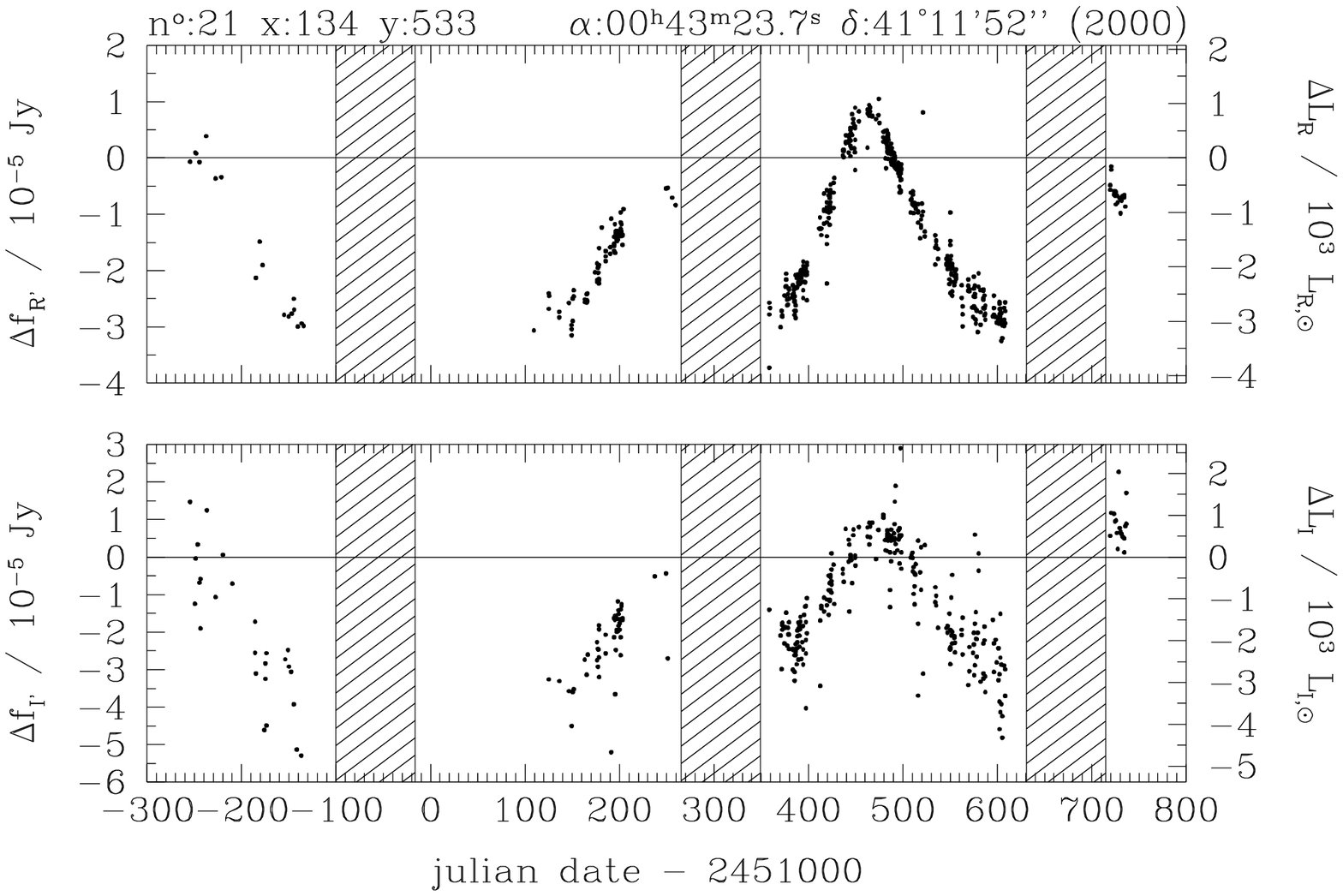}
\caption{Light curve of a longperiodic variable. 
Upper panel: $\mathrm{R'}$ band, lower panel: $\mathrm{I'}$ band.
\vspace*{.38cm}}
\label{lc_06}
\end{figure}

\begin{figure}
\centering
\includegraphics[width=0.48\textwidth]{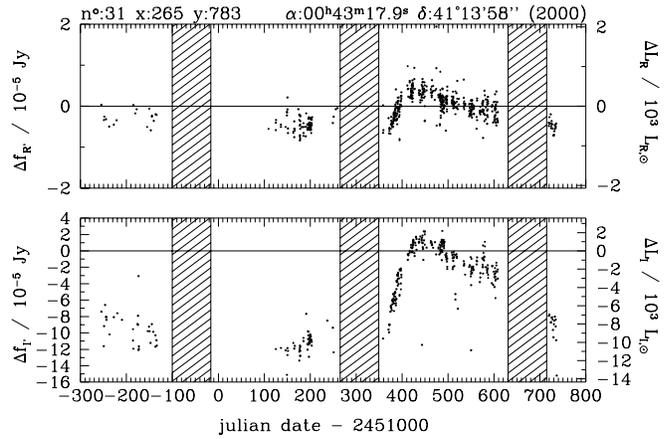}
\caption{Light curve of a longperiodic variable star with 
a very large variation in the $\mathrm{I'}$ band. Upper panel: 
$\mathrm{R'}$ band, 
lower panel: $\mathrm{I'}$ band.
\vspace*{.4cm}}
\label{lc_07}
\end{figure}

\begin{figure}
\centering
\includegraphics[width=0.48\textwidth]{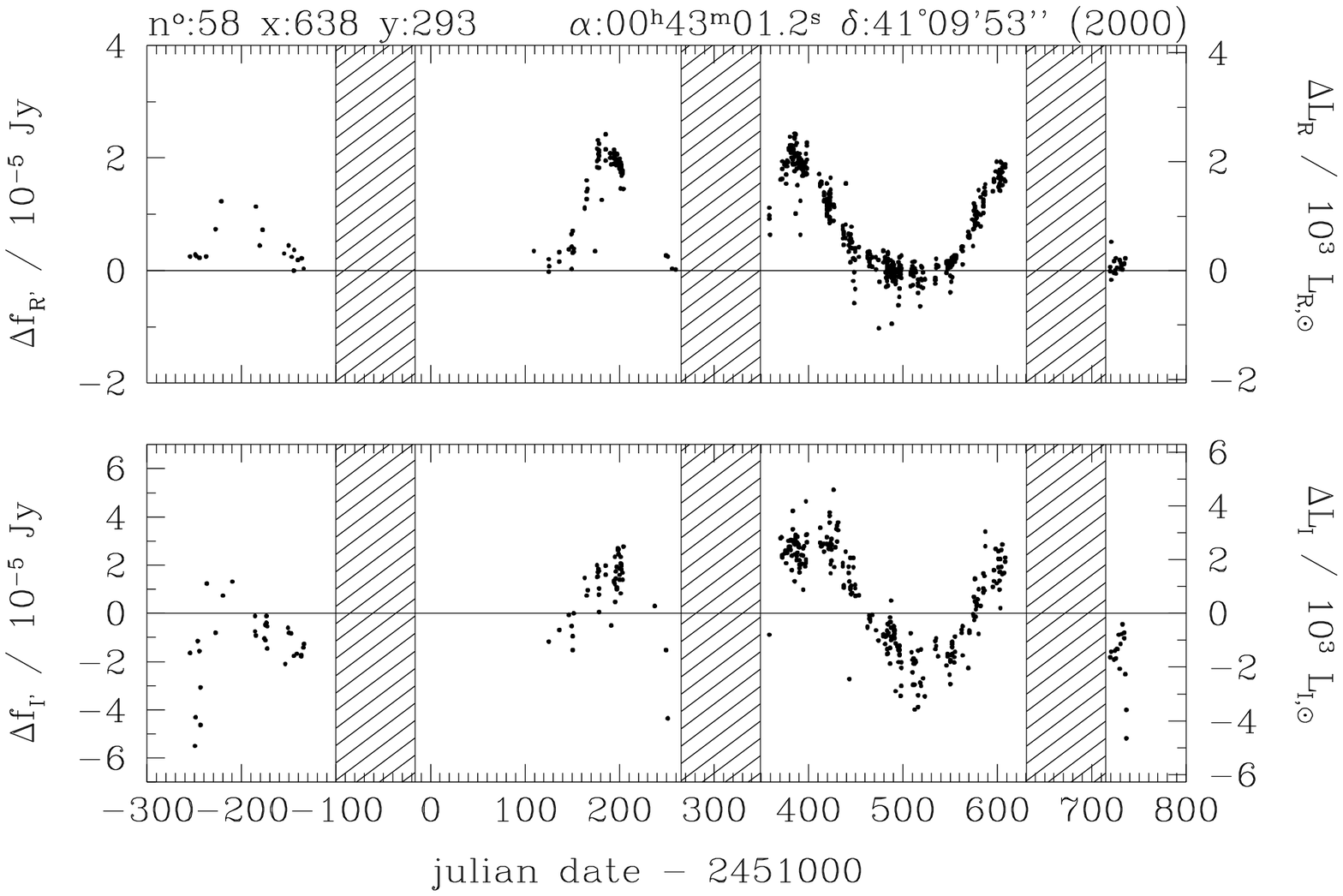}
\caption{Light curve of a longperiodic variable. 
Upper panel: $\mathrm{R'}$ band, lower panel: $\mathrm{I'}$ band.}
\label{lc_08} 
\end{figure}

\begin{figure}
\centering
\includegraphics[width=0.48\textwidth]{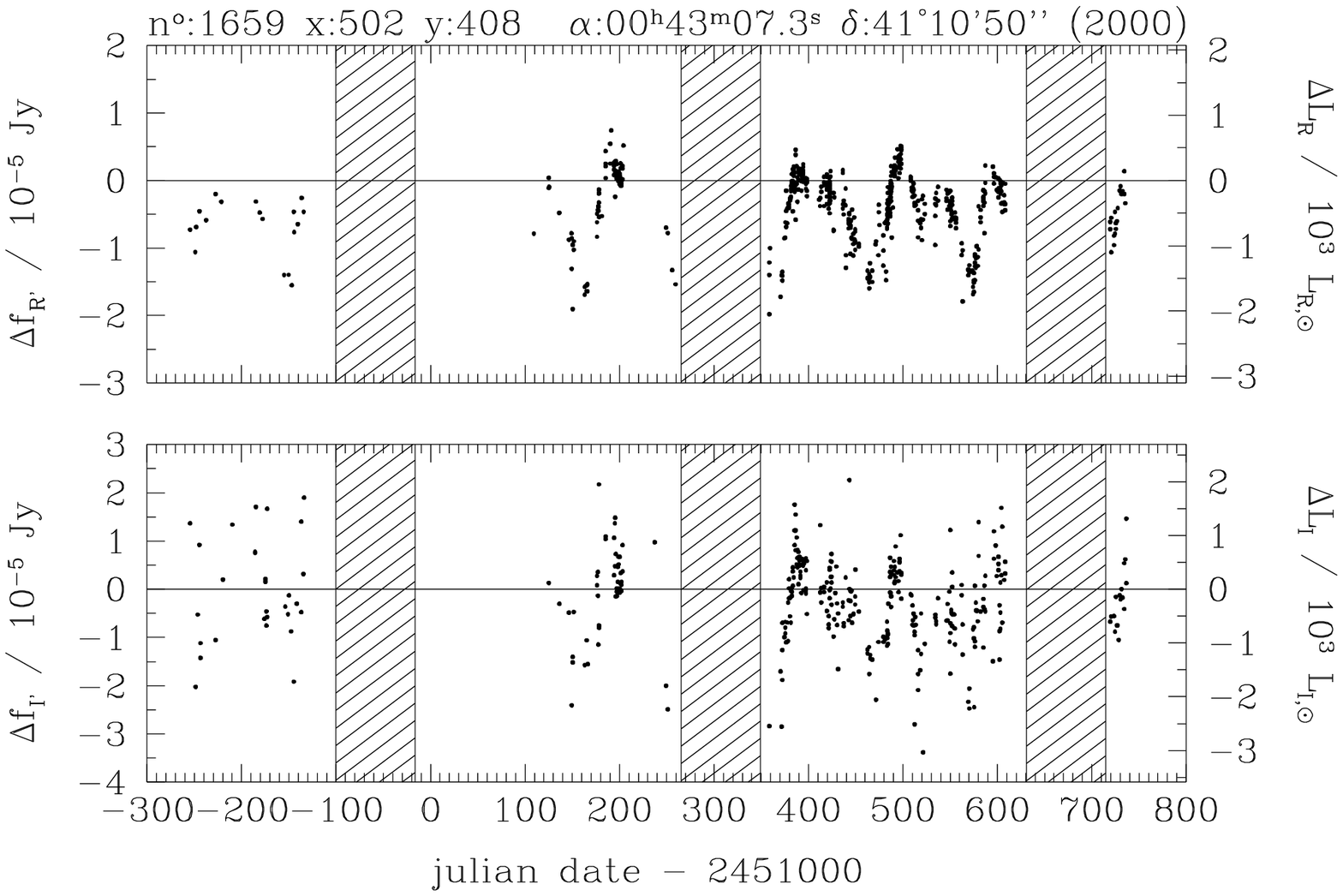}
\caption{Light curve of a RV Tauri star in the 
$\mathrm{R'}$ (upper panel) and $\mathrm{I'}$ (lower panel) bands. 
Due to the optimal
time coverage, the typical double-wave shape with alternating deep and 
shallow maxima of the light curves of this class of variable stars 
is uncovered.}
\label{lc_10}
\end{figure}

We present a small sample of light curves to show the efficiency of
the method. All light curves were observed over more than three years
from 1997 until 2000. Time spans when M\,31 was not observable
are marked by shaded regions. Because of bad dome seeing conditions 
and an inappropriate autoguiding system errors were largest during 
the first Wendelstein campaign 1997/98. During the second period 1998/99 
we were able to decrease the FWHM of the PSF by a factor of two, thus 
the photometric scatter is also clearly smaller. 
During the third period 1999/2000 we observed simultaneously at Calar Alto and
Wendelstein and got data points for 53\,\% of the visibility of M\,31.
\\
The OIS method can be applied for very crowded fields like M\,31 and gives 
residual errors at the photon noise level (Fig.~\ref{photon_noise}), 
see also \citet{2001AA..00..00}).
A good estimate for the average noise present in the area 
$\Omega_{\mathrm{PSF}}$ of a PSF is $N=0.1\times10^{-5}\mathrm{Jy}$.
The light curves of variable stars presented in the Figs.~\ref{lc_01} 
through \ref{lc_10} indicate a typical scatter which is in good agreement 
with the above estimate. This means that a red clump giant with a 
brightness of $M_\mathrm{I}=0$ \citep[Fig.~7]{1996AJ....112.1975G} and a 
colour of $(R-I)=0.5$ \citep{1998A&AS..130...65L} has to be 
amplified by a factor of 10 to be detected with a peak signal-to-noise 
ratio of $(S/N)=3$ in our survey. The brightest RGB stars
with a $M_\mathrm{I}=-3.5$ and a colour of $(R-I)=1$ need an 
amplification of 1.6
only.
\\
Up to now we detected over 5000 variable sources in a 
$8\times8\,\mathrm{arcmin}^2$ field. 
A preliminary analysis of the light curves shows that 
we have found the whole range of variable stars including novae and other 
types of eruptive variables, Cepheids, semi-regular, Mira-type and other 
longperiodic variables. 
\\
In Fig.~\ref{lc_01} we present one of the 
$\delta$-Cephei variable stars in the $\mathrm{R'}$ and $\mathrm{I'}$ bands, 
Fig.~\ref{convol1} 
shows the $\mathrm{R'}$ light curve of this star convolved with its period, 
which was determined to $15.76 \pm 0.01$ days.
Figure~\ref{lc_02} presents the light curve of a nova previously
published by \citet{1999IAUC.7218....2M}. It's the brightest variable 
source detected in our M\,31-field.
Figure~\ref{lc_09} is an example for an eruptive variable star, which could 
be mistaken as a microlensing event, if the time coverage were 
insufficient. 
Figures~\ref{lc_03} to \ref{lc_08} display light curves of variable stars, 
which were classified as longperiodic in a preliminary analysis. Finally 
we present the light curve of a RV Tauri star in Fig.~\ref{lc_10}.

\section{Conclusions}

We presented an overview of the Wendelstein Calar Alto Pixellensing Project
(WeCAPP). We demonstrated that despite observing at different sites with
different instruments all data can be used for optimal image
subtraction following \citet{1998ApJ...503..325A}. This method can be applied
for very crowded fields like M\,31 and gives residual errors at the
photon noise level. A red clump giant of $M_\mathrm{I}=0$, 
which is amplified by a factor of 10 by a microlensing event, 
can be detected with our data.
We showed how the data are reduced and how light curves are extracted. 
For illustration we presented a small sample of light curves. In future 
publications we will present a full catalogue of variable sources 
which we found in our M\,31 field, including potential MACHO light curves.

\begin{acknowledgements}
The authors would like to thank the staff at Calar Alto Observatory
for the extensive support during the observing runs of this project.
Special thanks go to the night assistants for all the service observations
carried out at the 1.23 m telescope: F. Hoyo (60\,\% of all service 
observations), S. Pedraz (20\,\%), M. Alises (10\,\%), A. Aguirre (10\,\%), 
J. Aceituno, and L. Montoya. 
The Calar Alto staff members H. Frahm, R. Gredel, F. Prada, and  U. Thiele
are  especially acknowledged for their instrumental and astronomical support.
\\
Many thanks go to W. Wimmer, who helped to improve the seeing
conditions at the Wendelstein telescope.
\\
We acknowledge stimulating discussions with N.~Drory, G. Feulner, A.~Fiedler,
A.~Gabasch,  and J.~Snigula.
This work was supported by the \emph{Sonder\-forschungs\-bereich, 
SFB 375}, Astro\-teil\-chen\-phy\-sik.
\end{acknowledgements}

\mail{arri@usm.uni-muenchen.de}

\end{document}